\documentclass[aps,preprintnumbers,showpacs,twocolumn,groupedaddress,floatfix,nofootinbib]{revtex4-1}

\usepackage{latexsym} 
\usepackage{amssymb} 
\usepackage{amsmath} 
\usepackage{amsfonts}
\usepackage{bm}
\usepackage{color}
\usepackage{times} 
\usepackage{units}
\usepackage{hyperref}

\usepackage[utf8x]{inputenc}
\usepackage{graphicx} 
\usepackage[squaren]{SIunits}

\newcommand{\Fref}[1]{Fig.~\ref{#1}} 
\newcommand{\Sref}[1]{Sec.~\ref{#1}} 
\newcommand{\Tref}[1]{Table~\ref{#1}}

\newcommand{\parsec}{\mathrm{pc}}

\begin{document}

\title{Structure of stable binary neutron star merger remnants: A case study}

\author{W. Kastaun, 
R. Ciolfi, 
B. Giacomazzo}

\affiliation{Physics Department, University of Trento, 
via Sommarive 14, I-38123 Trento, Italy}

\affiliation{INFN-TIFPA, Trento Institute for Fundamental Physics and Applications,\\
via Sommarive 14, I-38123 Trento, Italy}


\begin{abstract}
In this work, we study the merger of two neutron stars with a gravitational mass 
of $1.4\usk M_\odot$ each, employing the Shen-Horowitz-Teige equation of state.
This equation of state is a corner case, allowing the formation of a stable
neutron star with the given total baryonic mass of $3.03 \usk M_\odot$.
We investigate in unprecedented detail the structure of the remnant, in particular
the mass distribution, the thermal structure, and the rotation profile.
We also compute fluid trajectories both inside the remnant and those 
relevant for the formation of the disk.
We find a peanut-shaped fluid flow inside the remnant following a strong 
$m=2$ perturbation. Moreover, the flow is locally compressive, causing the 
appearance of dynamic hot spots. Further, we introduce new diagnostic measures
that are easy to implement in numeric simulations and that allow to quantify 
mass and compactness of merger remnants in a well-defined way. As in previous
studies of supra- and hypermassive stars, we find a remnant with a slowly rotating 
core and an outer envelope rotating at nearly Keplerian velocity. 
We compute a Tolman-Oppenheimer-Volkoff star model which agrees well with that of the remnant in the 
core, while the latter possesses extensive outer layers rotating close to Kepler 
velocity. Finally, we extract the gravitational wave signal and discuss the detectability
with modern observatories.
This study has implications for the interpretation of gravitational wave 
detections from the post-merger phase, and is relevant for short
gamma-ray burst models.
\end{abstract}


\maketitle

\section{Introduction}
\label{sec:intro}

Binary neutron star (BNS) mergers are one of the most interesting astrophysical
scenarios since their description requires both general relativity (GR) in the strong field
regime as well as nuclear physics for matter above nuclear density, in particular
the equation of state (EOS) and neutrino physics.
BNS mergers are also promising candidates for multimessenger astronomy.
First, they are among the most promising sources of gravitational waves 
for the detection with ground-based interferometers such as advanced LIGO \cite{Harry:2010:84006}
and Virgo \cite{Virgo2015}. Second, they are thought to be the energy source  
of short gamma ray bursts (SGRBs) \cite{Berger2014}, which are frequently observed 
among the most energetic explosions known in the universe.
Furthermore, these events represent the most likely explanation for the 
abundance of heavy elements in the universe, complementary to supernova 
explosions \cite{Thielemann2011, Wanajo2014}.
The hot and neutron-rich ejecta form the merger process can indeed undergo rapid neutron 
capture nuclear reactions (r-process reactions) which produce the heavy elements. 
Moreover, the radioactive decay of these elements heats up the material and leads to a 
so-called kilonova (or macronova) signal \cite{Li:1998:L59, Kulkarni:2005:macronova-term,
  Metzger2012}, emerging mostly in the optical band 
days/weeks after merger. Candidates for such a kilonova have already 
been found \cite{Tanvir2013, Berger2013}. Finally, in addition to SGRBs and kilonova signals, BNS mergers 
may be accompanied by other electromagnetic counterparts, including 
late-time radio signals (e.g., \cite{Metzger2012}) and, if the remnant is not 
a black hole (BH) but a long-lived metastable (or stable) NS~\cite{Giacomazzo2013ApJ...771L..26G}, luminous signals
powered by the remnant spin-down and peaking in the x-ray band 
\cite{Siegel:2016a,Siegel:2016b,Fan:2006:372L19} or at lower energies \cite{Yu2013,MetzgerPiro2014}.

The recent direct detection of two gravitational wave (GW) events
caused by binary BH mergers \cite{LIGO:BBHGW:2016,LIGO_GW151226} 
has raised expectations for a detection of GWs from BNS mergers in the near future 
\cite{Abadie2010}.
Such a detection might include a relatively strong signal from the post-merger phase.
While the post-merger signal of binary BH mergers always consists of a short 
BH ringdown, for a BNS merger it can be much more complex. Moreover, this signal 
offers a unique opportunity to constrain the NS EOS~\cite{Bauswein:2012:11101,
  Takami:2014:91104, Takami2015}.  

The outcome of a BNS merger depends largely on the total mass of the system:
while heavy systems promptly collapse to a BH during merger, lighter systems form
a metastable or even stable NS. 
Merger remnants can be classified based on their mass. A hypermassive NS has a mass
exceeding the maximum mass of a uniform rotating star, and a supramassive NS is 
below that mass but above the maximum mass for a non-rotating configuration.
Those masses are also used to predict the lifetime of the merger remnant.
A hypermassive neutron star (HMNS) collapses within tens of milliseconds, as soon as
the degree of differential rotation becomes insufficient to stabilize it.
Above a certain threshold mass, however, the merger results in a prompt collapse without a 
HMNS phase \cite{Bauswein:2013:131101}.
A supramassive neutron star (SMNS) can be supported by uniform rotation for much longer 
spindown timescales, and eventually collapses when the rotational support becomes too small. 
In the early post-merger phase ($<100$ ms) the associated loss of angular momentum 
can be dominated by GW emission, while magnetic braking likely dominates on longer timescales.

The exact values of the masses dividing the above cases depend sensitively on the EOS.
Hence, the presence of a post-merger signal could rule out all EOS for which
the masses determined from the inspiral signal would lead to a prompt collapse.
Moreover, the detection of a very long post-merger signal (although more difficult due to decaying
amplitude), could even rule out EOS predicting a HMNS as the outcome, instead of a SMNS.
In addition, the post-merger GW spectrum is typically dominated by a single peak around a few 
kilohertz, which also depends on the EOS. Measuring the corresponding frequency might allow 
for an additional constraint on the EOS 
\cite{Bauswein:2012:11101,Takami:2014:91104, Takami2015}.  

In addition to the post-merger GW signal, there are additional reasons to consider a BNS merger
leading to a long-lived remnant NS as a very important case.  
First, the recent discovery of single NSs with a mass of $\sim2~M_\odot$ 
\cite{Demorest:2010:1081,Antoniadis:2013:448} combined with the fact that the expected 
distribution of progenitor masses in BNS mergers is peaked around $1.3-1.4~M_\odot$
\cite{Belczynski2008} leads to the conclusion that forming a SMNS is a likely possibility, 
and in some cases it might be possible to form even a stable NS 
\cite{Giacomazzo2013ApJ...771L..26G}.

A second reason is that recent observations of SGRBs by Swift \cite{Gehrels2004} 
revealed a large fraction of events accompanied by long-lasting (from minutes to hours) 
x-ray afterglows, which suggests the presence of a continuous injection of energy from a 
long-lived central engine (e.g.,~\cite{Rowlinson2013}).
If SGRBs are associated with BNS mergers, this would suggest the presence of a 
long-lived (supramassive or stable) NS, instead of the commonly assumed BH surrounded 
by an accretion disk.
The main difficulty of this so-called ``magnetar model" is the fact that in this case 
the strong baryon pollution surrounding the merger site could choke the formation 
of a relativistic jet and thus prevent the SGRB itself. To overcome such limitations, 
an alternative ``time-reversal" scenario \cite{Ciolfi:2015:36} has been proposed, in which 
a SMNS survives for some time powering the long-lasting x-ray emission and eventually 
collapses to a BH and launches the relativistic jet. Because of the optical depth of the surrounding 
environment, the spin-down-powered x-ray emission is delayed and can still be observed 
after the SGRB (thus appearing as an ``afterglow").
The main challenge to this scenario is the formation of a massive disk surrounding the BH, 
when the latter is formed not shortly after merger, but minutes or hours later. 
Studying the rotation profile, the disk structure and the amount of ejecta 
surrounding the merger site in the case of a long-lived NS is therefore necessary in order to 
shed light on the different SGRB scenarios.

Models of differential 
rotation in the remnant NS often assume a rapidly rotating core, with the rotation profile described
by the so-called $j$-const law \cite{Baumgarte:2000:29}. 
In contrast, a recent study \cite{Kastaun:2015:064027} demonstrated 
that HMNS or SMNS can indeed be produced with a slowly rotating core. In the case of HMNSs, 
the collapse can be significantly delayed because part of the mass forms 
an extensive outer bulge rotating close to Kepler velocity, thus centrifugally supported. 
Further studies \cite{Endrizzi:2016:164001} found similar behavior for more models, including
an unequal mass system. 
The rotation profile is important for the oscillation frequencies of the core and
hence the radiated GW signals. Also, the lifetime of a HMNS stabilized by
a rapidly rotating core could differ strongly from a HMNS stabilized by a rapidly 
rotating outer bulge. This should be taken into account when attempting to deduce the
EOS from the duration of a postmerger GW signal. 
Moreover, the differential rotation profile may also have an effect on
magnetic field amplification and electromagnetic emission, as seen in
previous studies employing the $j$-const
law~\cite{Siegel:2014:6}. Therefore a careful investigation of the
remnant properties also has important applications to electromagnetic
counterparts of GWs from BNS mergers and, possibly, to SGRBs.

Beside rapid differential rotation and larger mass, another difference to isolated NS
is that shock heating during merger requires the consideration of thermal effects.
A common notion is that HMNS are also stabilized by thermal pressure. For nuclear physics 
EOS however,
the temperatures reached during merger do not increase the maximum mass of spherical or
uniformly rotating NSs strongly, and, somewhat counter-intuitively, can even decrease 
it \cite{Kaplan:2014:19,Galeazzi:2013:64009}. However, the merger remnants are not heated 
homogeneously. It is possible
that under-densities caused by ``hot spots'' in the rapidly rotating remnant contribute 
to the GW signal, with a frequency determined by the rotation rate at the location
of the hot spot. Unless the hot spots occur near a maximum of the rotation rate,
they would quickly turn into tightly wound spirals because of differential rotation,
at which point the associated GW signal ceases. Because of this effect, both the exact 
nature of the fluid flow inside the remnant and the thermal structure, as well as
their interplay, could become important for the interpretation of future GW detections.
Note however that such contributions are likely smaller than the one from the non-axisymmetric
oscillation of the remnant. Further, this effect competes with other secondary features
caused by nonlinear combination frequencies between the radial and the main non-axisymmetric 
oscillation \cite{Stergioulas:2011:427} or side peaks caused by a strong frequency modulation
of the main peak \cite{Kastaun:2015:064027}, again caused by radial oscillations.
Even with a good signal to noise ratio for the post-merger phase, deriving information from 
secondary features of the post-merger spectrum will require a good theoretical understanding 
of all those effects.

In this work, we will focus on a binary with typical mass but an EOS that allows the formation
of a stable NS. 
This case is generically considered rather unlikely and for this reason it has 
not been investigated in detail. We note, however, that most of the results that will be 
presented in this study are also indicative for the SMNS case, which is instead a likely 
outcome of a BNS merger. 
We will investigate the structure of the fluid flow and thermal structure 
with regard to possible contributions to the post-merger GW signal. We will also investigate the rotation profile
and mass distribution, and then we will take a closer look at the formation of the disk
surrounding the remnant. Finally, we provide estimates for the mass ejection.

Throughout this paper, we use units $G=c=1$ unless noted otherwise. We define baryonic mass
$M_b = N_b m_b$ and rest mass density $\rho = n_b m_b$, where $N_b$ and $n_b$ are baryon number and
baryon number density in the fluids restframe, respectively, and $m_b$ is a formal mass constant
$m_b = 931.494 \usk\mega\electronvolt$.

\section{Setup}
\label{sec:numerics}

\subsection{Models}
\label{sec:models}

In this study, we consider an irrotational binary neutron star system consisting of
two stars with a gravitational mass (in isolation) of $1.4\usk M_\odot$ each.
We use the Shen-HorowitzTeige (SHT) EOS
\cite{Shen:2010:15806,Shen:2011:35802}, which is 
based on a relativistic mean field theory method and a modified NL3 set 
of interaction parameters. The initial model is at zero temperature and 
in $\beta$-equilibrium.

The total baryonic mass of the binary is  $3.03\usk M_\odot$. For such a high mass, the 
outcome of the merger would be a HMNS or a prompt collapse for most EOS.
The SHT EOS, however, features an unusually large maximum mass of
$3.33\usk M_\odot$ for Tolman-Oppenheimer-Volkoff (TOV) stars. The maximum mass for uniformly rotating
stars is $3.97\usk M_\odot$. Therefore, we expect that a stable NS is formed 
after merger. 

The initial data were computed using the \texttt{LORENE} code 
\cite{Gourgoulhon:2001:64029}. Note that because the quasicircular approximation
employed in the formalism ignores the radial velocity component of the inspiral,
the resulting models always show a small eccentricity when evolved.
We did not attempt to correct for this, as done, e.g., in~\cite{Dietrich:2015:124007}, since the 
eccentricity is mainly a problem when studying tidal effects during the inspiral.
The initial proper separation of our model is $57.6 \usk\kilo\meter$, which corresponds
to 4 orbits until merger.

\subsection{Numerical methods}
\label{sec:code}

To evolve the general relativistic hydrodynamic equations, we use the WhiskyThermal 
code described in \cite{Galeazzi:2013:64009,Alic:2013:64049}. It is based on a 
finite-volume high resolution shock capturing scheme that uses the Harten-Lax-van Leer-Einfeldt approximate
Riemann solver and the piecewise parabolic method for reconstructing values at the 
cell interfaces. We employ a tabulated equation of state including thermal and 
composition effects in terms of the electron fraction. We do not consider magnetic
fields and neutrino radiation, and the electron fraction is passively advected along the
fluid.

Our code features a robust
method (see \cite{Galeazzi:2013:64009}) for the conversion from evolved to primitive 
variables that also includes stringent checks of the evolved variables, with a clear 
error policy, allowing for adjustments of harmless numerical errors such as temperatures 
falling slightly below zero when evolving cold initial data, but aborting the run in case 
of severe errors. The error policy used here is the same as described in 
\cite{Galeazzi:2013:64009}.

For this work, we also included the option to enforce adiabatic evolution at zero
temperature. The motivation is to avoid artificial heating at the NS surfaces during 
the inspiral, which is a cumulative effect. To enforce zero-temperature evolution,
we evolve in the same way as before, including the energy, but recompute the conserved
energy after each evolution sub-step from conserved density and momentum, assuming zero 
temperature. For this, we modified our conservative to primitive scheme by replacing the 
specific energy computed from the evolved variables by the value obtained from the EOS 
evaluated at zero temperature (more precisely, at the lowest tabulated value, which is 
low enough for our purpose). This way, the evolved conserved energy is not used at all. 
Around a $\milli\second$ before the stars touch, we activate the full thermal evolution.

For the evolution of the spacetime, we use the \texttt{McLachlan} code \cite{Brown:2009:44023},
which is part of the Einstein Toolkit \cite{Loeffler:2012:115001}. This code implements two 
formulations of the evolution equation. Instead of the popular Baumgarte-Shibata-Shapiro-Nakamura formulation 
\cite{Nakamura:1987:1,Shibata:1995:5428,Baumgarte:1998:24007}, 
we chose the newer conformal and spatially covariant CCZ4 evolution scheme described 
in \cite{Alic:2012:64040, Alic:2013:64049}, which has constraint damping capabilities.
The gauge conditions used during evolution are the $1+\log$-slicing condition 
\cite{Bona:1995:600} for the lapse function and the hyperbolic $\Gamma$-driver
condition \cite{Alcubierre:2003:84023} for the shift vector. At the outer boundary,
we use the Sommerfeld radiation boundary condition.
 
All codes are integrated into the Cactus Computational Toolkit infrastructure.
The time evolution is coupled using the method of lines, in particular we use
a fourth order Runge-Kutta time integrator. Further, we make use of Berger-Oliger
moving-box mesh refinement provided by the \texttt{Carpet} code \cite{Schnetter:2004:1465}.
In detail, we use six refinement levels. The two finest ones follow the stars during
inspiral, and are replaced by fixed levels centered around the origin shortly before merger.
The finest grid spacing is $295 \usk\meter$. The outer boundary is located at $945\usk\kilo\meter$,
and the finest level after merger covers a radius of $30\usk\kilo\meter$.
Finally, we use reflection symmetry across the orbital plane.
For tests of the code, we refer the reader to \cite{Galeazzi:2013:64009,Alic:2013:64049}.

\subsection{Analysis tools}
\label{sec:tools}

Since we are interested in the perturbation of the merger remnant, we need to be able 
to distinguish between actual deformations and mere gauge effects. In particular the 
spatial coordinates in GR simulations can be problematic for the analysis of deformations, 
since the gauge conditions typically used are mainly designed to prevent coordinate 
pathologies, not to seek symmetries. Instead of changing the gauge conditions during 
the simulation, we transform to a more suitable coordinate system during the 
post-processing stage. The new coordinates are obtained following a well-defined
prescription we described in detail in \cite{Kastaun:2015:064027}. 

Those coordinates
are restricted to the orbital plane (and require reflection symmetry). The main 
advantages are that the radial coordinate is parametrized by proper length, 
that the radial coordinates are on average orthogonal to the $\phi$-coordinates, 
preventing spiral deformations, and, most important,
that the $\phi$-coordinate base vector becomes a Killing vector for an axisymmetric 
spacetime.

To measure deformations of the remnant, we use a decomposition in $\phi$-harmonic
components as described in \cite{Kastaun:2015:064027}:
\begin{align}
P_m^q &= \int_0^R  \int_0^{2\pi} q(r,\phi) e^{i m \phi} \mathrm{d}A
\end{align}
where $q$ stands for an arbitrary scalar variable, $\mathrm{d}A$ is the proper area element
in the new coordinates $(r,\phi)$, and $R$ is a suitable cutoff radius larger than the 
remnant. We are mostly interested in $P^\rho$, for which we use $R=40\usk\kilo\meter$,
large enough to fully contain the remnant.

Besides oscillations of the remnant, we also want to measure its radial structure.
This is, however, complicated by the fact that merger remnants are rapidly and differentially
rotating objects, which typically also exhibit strong non-axisymmetric oscillations
excited during the merger. Together with the ambiguity of defining radii in GR, it becomes
non-trivial to define, e.g., the density profile in the same way as for a spherically
symmetric static star. To solve this issue we consider, for a given time slice, 
surfaces of constant baryon mass density (baryon number density), measured with respect 
to the fluid rest frame. Those only depend on the chosen foliation of spacetime, but not on
any spatial coordinate system. For each density isosurface, we can measure the total enclosed
proper 3-volume $V$, where the enclosed region is defined as the region occupied by matter with 
higher densities. In the same way, we define the total enclosed baryon mass $M_b$.
This way, we have established a mass-versus-volume relation that is a suitable replacement
for density-versus-radius relations used to describe spherical stars. For convenience,
we also define a volumetric radius $R_V$ as $V = \frac{4}{3}\pi R_V^3$.

Note that although this measure is mainly intended for objects of spherical topology, it 
is well defined for any density distribution.
Examples with the topology of a torus are given by the disk around a merger remnant, as
well as the high-density regions of some hypermassive NS models featuring a ring-shaped 
density maximum. For such cases, the volumetric radius cannot be interpreted as a measure 
of the object size, but the mass-volume relations can still be used to make unambiguous 
comparisons between such distributions.

Another important measure of spherical stars is the compactness $M/R$, usually defined
in terms of gravitational mass and circumferential radius. Since both are hard to define
unambiguously for merger remnants in GR, we use the baryon mass and the volumetric radius
instead to define the compactness $C_V = M_b/R_V$ of a given isodensity surface. Defining 
the compactness
of the remnant, we encounter another obstacle given by the fact that instead of a NS surface
there is a smooth transition to the surrounding disk and other merger debris. To circumvent 
this, we use the observation that our new compactness measure $C_V$ has a clear maximum.
We define the ``bulk'' as the region enclosed by the isodensity surface with the
maximum compactness, and the enclosed baryon mass and proper volume as bulk mass 
$M_\mathrm{blk}$ and bulk volume $V_\mathrm{blk}$. Note that $C_V$ is designed for use 
with isolated stars and merger remnants, where the bulk will most likely always be a 
moderately deformed sphere. Although still well defined when applied to a torus for example, 
it could not be interpreted any more as mass within some average radius.

In order to motivate the tentative name choice bulk, we now derive some expressions
for the density and density gradient on the bulk surface.
For this discussion, we assume that the isodensity surfaces can be used as a smooth 
foliation of space, and express the integrals for enclosed volume and mass as
\begin{align}
V(\rho') &= \int_{\rho'}^\infty \int_{A(\rho)} \mathrm{d}V,\\
M_b(\rho') &= \int_{\rho'}^\infty \int_{A(\rho)} W\rho\, \mathrm{d}V,
\end{align}
where $\mathrm{d}V$ is the proper 3-volume element, $A(\rho)$ the isodensity surface
of rest-mass density $\rho$, and $W$ the fluid Lorentz factor. Decomposing the volume
element as $\mathrm{d}V = N \,\mathrm{d}A \,\mathrm{d}\rho$, 
where $\mathrm{d}A$ is the proper area element on $A(\rho)$, and $N$ is a scalar 
function (analogous to the lapse function of space-time foliations), we find
\begin{align}
\frac{dM_b}{dV} &= \rho \bar{W}(\rho), &
\bar{W}(\rho) &= \frac{\int_{A(\rho)} W N \mathrm{d}A}{\int_{A(\rho)} N \mathrm{d}A},
\end{align}
and, together with the definition of $C_V$,
\begin{align}
\frac{dC_V}{dV} &= V^{-\frac{1}{3}} \left( \rho \bar{W}(\rho) -\frac{M_b}{3V} \right).
\end{align}
Thus, a surface of maximum compactness has to satisfy
$3\bar{W}(\rho_\text{blk}) \rho_\text{blk}  = M_\text{blk}/V_\text{blk}$. 
In other words, the density of the bulk surface (times its average Lorentz factor $\bar{W}$) is 
one third of the average interior density $M_b/V$.

Evaluating the second derivative of $C_V$ on the maximum compactness surface, 
we find
\begin{align}
0 &> \frac{d^2C_V}{dV^2} 
= V^{-\frac{4}{3}} \left( \frac{2}{3} \bar{W}\rho +V \frac{d}{dV}\left(\bar{W}\rho\right)\right).
\end{align}
Inserting the definition of $R_V$, we obtain
\begin{align}
\frac{d\ln\left(\bar{W}\rho\right)}{d\ln R_V} &< -2.
\end{align}
Hence, $\bar{W}\rho$ falls of faster than $R_V^{-2}$ near the bulk surface.

When applied to our merger simulations, the definition of the bulk excludes the 
outermost low-density layers of the remnant as well as the disk.
We note that the bulk properties are influenced by rapid rotation and nonradial 
oscillations only via changes of average density, not by the deformation of the shape. In 
the limit of a constant density star and flat space, the bulk properties are even 
completely independent of its shape.

Besides baryon number and proper volume, one can also compute the 
total entropy $S_\mathrm{blk}$ as well as the number of electrons inside the bulk, 
defining the bulk average specific entropy $\bar{s}_\mathrm{blk} = S_\mathrm{blk} / M_\mathrm{blk}$ 
and bulk average electron fraction. The former will be used in this work as a measure of heating 
during merger, and the latter might be useful in simulations involving neutrino cooling.

Of course, our new measures can also be applied to TOV solutions for spherical NSs. 
Compared to the surface radius, the bulk volume has the advantage that it is
insensitive to the low density part of the star, and might thus be more useful for any
computation that relates the dynamics of the whole star to the compactness, e.g.
empirical estimates for BNS mergers. The bulk compactness might also be a candidate 
for universal relations of NS oscillation frequencies, replacing e.g. the effective 
compactness defined from the moment of inertia (compare \cite{Chirenti:2015:044034}).

The above measures are surprisingly easy to implement in a numerical simulation.
We simply collect the proper volume and baryon mass of all grid cells into a 
onedimensional histogram where the bins are given by the logarithm of the rest 
mass density. The only complication is to make sure that points covered by several 
refinement levels are only counted once. The required masks are the same required 
for standard reduction operations, such as the average, and are therefore readily 
available. To get, e.g., the total mass inside a given isodensity surface, we then 
simply compute the reverse cumulative histogram to obtain the total mass of all 
cells with densities higher than the given one.
Extracting the bulk then only requires a simple one-dimensional maximum search. 
From the histograms, one can also extract the isodensity surface which contains a 
given total baryon mass. This will be used to visualize the mass distribution.

In order to better understand the evolution during and after merger, we will study
the trajectories of fluid elements in addition to the above measures. The usual method 
to obtain fluid trajectories is to evolve tracer
particles during the simulation. The drawback of this approach is that the starting
positions have to be chosen before the simulation, and it is difficult to obtain a
good coverage of the computational domain at later stages. This is because the matter
in the disk surrounding the remnant is the result of an expanding fluid flow,
resulting in a tracer density which is much smaller than the original one.
Using a large number of tracers is inefficient, and guessing where to increase the initial 
tracer density is error prone.

To circumvent those problems, we evolve the fluid trajectories \emph{backward}
in time in a post-processing step. As starting positions, we simply use a regular
grid near the end of the simulation (or at whatever time we wish a homogeneous coverage).
This approach has the drawback that the data needs to be saved with a high enough
sampling rate to integrate the fluid trajectories with reasonable accuracy.
We therefore restrict ourselves to the equatorial plane, which also has the
advantage that we can use the well-defined coordinate system described above.
In addition, we integrate the trajectories in polar coordinates which are better 
suited for following the rapid rotation of the remnant than Cartesian coordinates.
Note also that when integrating backwards in time, the problematic expanding 
fluid flow becomes converging instead; thus an error in the integration is diminished
over time.

Last but not least, we need to determine the amount of matter ejected to infinity 
during the merger. As described in \cite{Kastaun:2015:064027}, this is a nontrivial
problem mainly due to the use of a constant-density artificial atmosphere in our 
code and because of the requirement to predict if a fluid element will eventually 
escape to infinity. We adopt the same tools described in previous works 
\cite{Endrizzi:2016:164001,Kastaun:2015:064027}. In short, we use the assumption of
geodesic motion as predictor for unbound matter and then compute the flux of unbound
matter through spherical surfaces of various radii as well as the volume integral
over the unbound mass density. Further, we collect the unbound mass in bins of 
coordinate radius at regular time intervals to get a spacetime diagram of the 
mass ejection.

\section{Results}
\label{sec:results}

\begin{figure*}
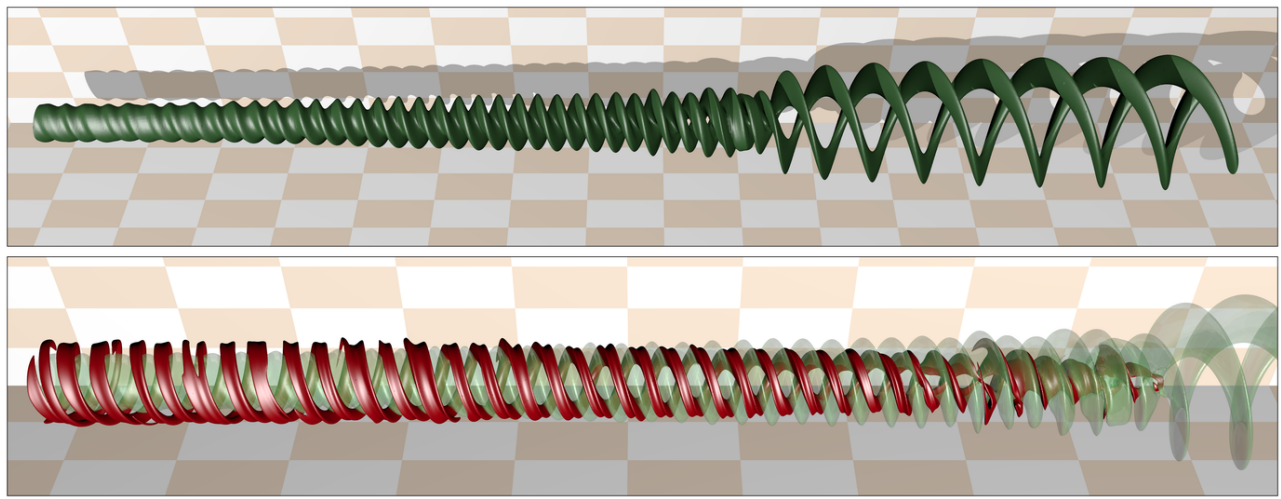

  \begin{center}
    \includegraphics[width=0.95\textwidth]{{{sht_s1_screwed_small}}}  
    \caption{Spacetime diagram showing the world tubes of different isocontours 
    in the orbital plane. The time coordinate runs from right to left.
    Top panel: isodensity contour containing (at each time) 25\% of the 
    total baryon mass, showing the evolution of the core(s).
    Bottom panel: the red surface is the contour of constant entropy density 
    chosen to highlight the evolution of the hot spots. For comparison,
    the surface in the top panel is shown again, but rendered transparent. 
    The size of the tiles in the background is 
    $2\usk\milli\second \times 20 \usk\kilo\meter$, and the lower panel is 
    zoomed to show the postmerger phase.}
    \label{fig:sht_s1_screwed}
  \end{center}
\end{figure*}

A visual overview of the evolution is given in \Fref{fig:sht_s1_screwed}, showing
spacetime diagrams where two dimensions correspond to the equatorial plane and the 
third to time.
The top panel depicts the evolution of the NS core(s).
To obtain this figure, we use the histograms introduced in \Sref{sec:tools}
to compute for each time the function mapping a given density to the total mass of all 
matter which has higher densities. We then map the density in space and time to the 
fraction of this mass to the total mass. The plot shows the isosurface that contains 
one quarter of the mass. To reduce gauge effects, we use the special coordinates 
defined in \Sref{sec:tools} for the orbital plane.

As one can see, the system completes 4 orbits before merging. At merger, the system undergoes
one ``bounce,'' after which it briefly consists of two separate cores in a common envelope.
Around $4\usk\milli\second$ after merger, a single strongly deformed remnant has formed.
Initially the deformation is dominated by the $m=2$-component. After $15\usk\milli\second$,
its amplitude is markedly decreased, and the influence of other components becomes visible.
This will be discussed in detail in \Sref{sec:deformation}.

To visualize the thermal evolution, we show the worldtube of the isocontours of entropy 
density in the equatorial plane in the lower panel of \Fref{fig:sht_s1_screwed}. 
We use the entropy density instead of temperature in order to focus on the remnant instead
of the surrounding matter.
A prominent feature visible in the figure is a two-armed helical structure in the post-merger
phase, corresponding to two entropy concentrations in circular motion. These hot spots
will be discussed in more detail in \Sref{sec:deformation}. Also note that during the 
double core phase, the entropy is concentrated around the origin. This is exactly what
is expected from shock heating during merger. In the subsequent evolution, the entropy 
density pattern apparently changes very quickly.

\begin{figure*}
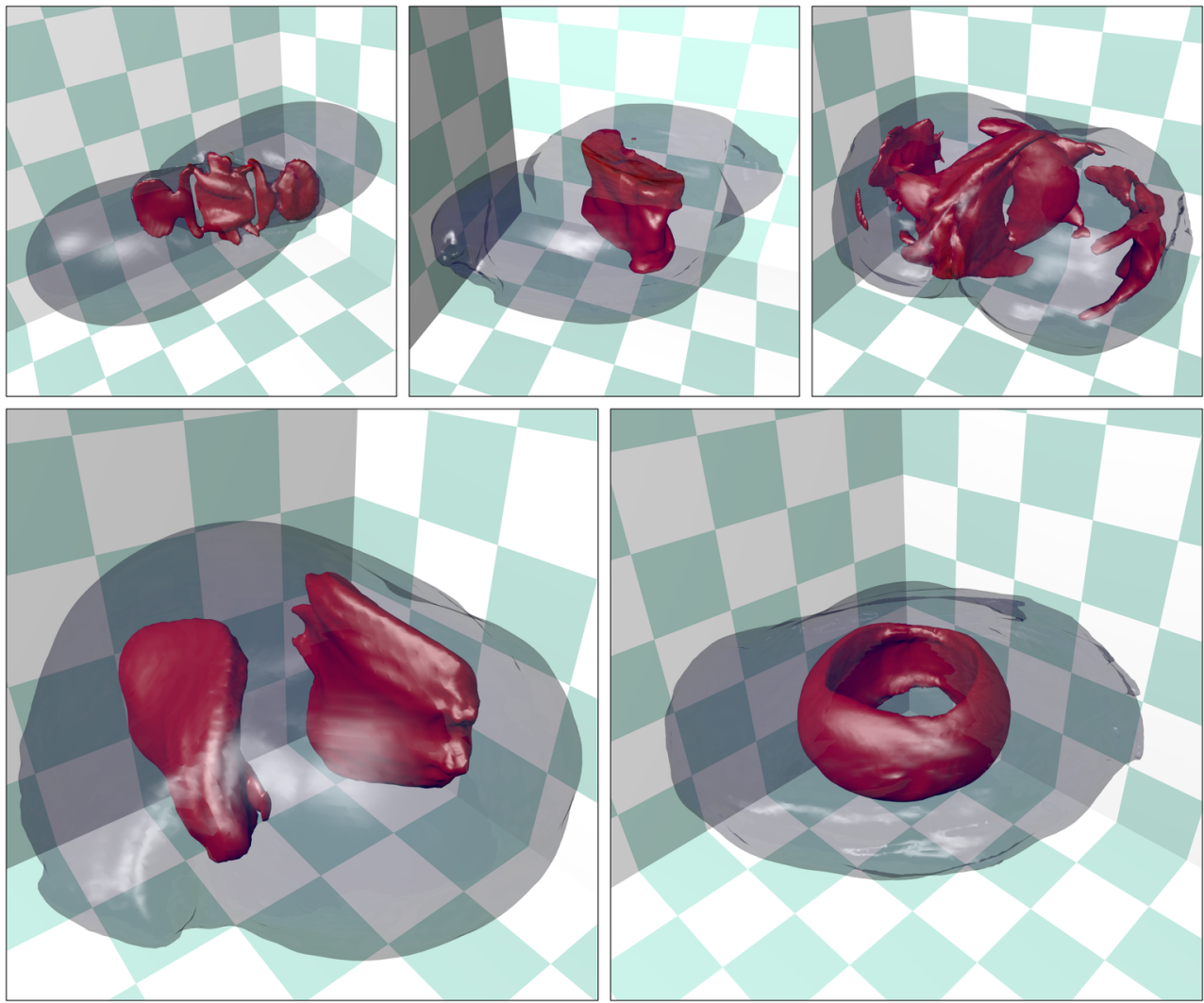

  \begin{center}
    \includegraphics[width=0.95\textwidth]{{{entropy_3d_snaps_small}}}  
    \caption{Snapshots of entropy distribution at times (from top-left to bottom-right) 
    $0.1$, $1.8$, $3.5$, $8.5$, and $18.3 \usk\milli\second$ after merger.
    The red surface is the isosurface of entropy density which contains 85\% of 
    the entropy. The transparent surface depicts the isosurface of mass density
    which contains a mass of $2.7\usk M_\odot$. The size of the tiles 
    in the background is $10\usk\kilo\meter$. }
    \label{fig:entropy_3d_snaps}
  \end{center}
\end{figure*}

The three-dimensional structure of the entropy distribution in the different evolution 
phases is shown in \Fref{fig:entropy_3d_snaps}. The first snapshot shows the shock-heated 
region produced when the stars touch, featuring three different parts. Apparently,
the ``elephant ears'' are produced  because the stars touch off-center in two points,
while the middle part is heated when the stars come in full contact.
The second snapshot shows the fully developed shock-heated region which separates the 
aforementioned double cores. This region is quickly ripped into different parts by the
complex fluid flow, as shown in the next snapshot. Shortly thereafter, the entropy is
concentrated in the two hot spots discussed above. Those entropy concentrations exhibit 
a saddle shape shown in the lower left panel. This pattern stays stable for several 
milliseconds. Afterwards, the system becomes gradually more axisymmetric until it 
assumes a ring-shaped configuration shown in the lower right panel. The snapshots also
show an isodensity surface containing 85\% of the mass.
This surface is strongly deformed up to the end of the simulation. In particular 
it exhibits a feature strongly resembling two wave-crests, one of which is visible 
in the lower left part of the panel.
By comparison with two-dimensional cuts, we found that those wave-crests are the origin of two
spiral waves that heat up the surrounding disk.

\begin{figure}
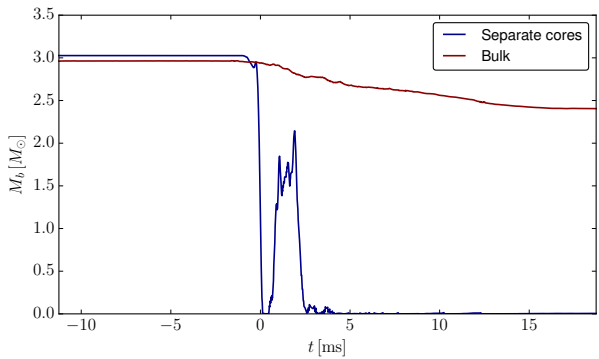

  \begin{center}
    \includegraphics[width=0.95\columnwidth]{{{core_masses}}}  
    \caption{Evolution of baryonic mass contained in the bulk and separate cores.
    The latter is defined as the mass of all matter with density higher than the  
    one at the origin.}
    \label{fig:core_masses}
  \end{center}
\end{figure}

The evolution of the bulk mass defined in \Sref{sec:tools} is shown in
\Fref{fig:core_masses}. After merger, it smoothly decreases from the value for the two 
isolated  stars by ${\approx} 0.5 \usk M_\odot$. This is due to the fact that
a significant fraction of the material is spread out in a rapidly rotating outer 
envelope and the disk. 
\Fref{fig:core_masses} also shows the total mass of matter with densities 
higher than the one at the origin. This measure is useful to quantify the 
double-core phase: the mass in the separate cores defined in this way is around 
$1.5$--$2\usk M_\odot$, and the length of the double core phase 
is~${\approx}2\usk\milli\second$.

\begin{figure}
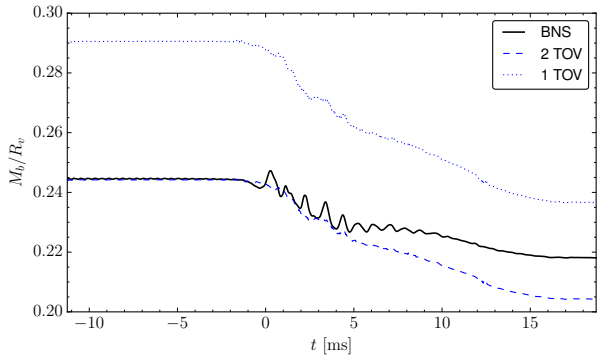

  \begin{center}
    \includegraphics[width=0.95\columnwidth]{{{bulk_comp}}}  
    \caption{Evolution of bulk compactness $M_b/R_V$ (solid line), where $R_V=((3/(4\pi))V)^{1/3}$ 
    is the volumetric radius of the bulk (see main text), $V$ its proper volume, and
    $M_b$ its baryon mass. For comparison, we show the same measure for a configuration with 
    the same bulk mass, but consisting either of a single TOV star (dotted line), or of 
    2 TOV stars (dashed line). The EOS for the latter is the same as for the initial data.}
    \label{fig:bulk_comp}
  \end{center}
\end{figure}

To quantify the compactness of the remnant, we use the bulk compactness defined
in \Sref{sec:tools}, which is shown in \Fref{fig:bulk_comp}.
For comparison, we also computed at each time the same 
measure for a single (cold) TOV star with the same bulk mass as the evolved 
system, as well as for a configuration of two isolated cold TOV stars.
Note the bulk compactness of each single star alone is (by definition) lower 
by a factor of $2^\frac{2}{3}$. We find that the bulk compactness is very close to the 
one for two isolated stars right until the merger. Apart from some oscillations, 
it starts deviating significantly only after the double core phase. Not surprisingly, 
the bulk of the rotating remnant is less compact than a TOV star with the same mass.
We will investigate the mass distribution in more detail in \Sref{sec:remnant}.

\begin{figure}
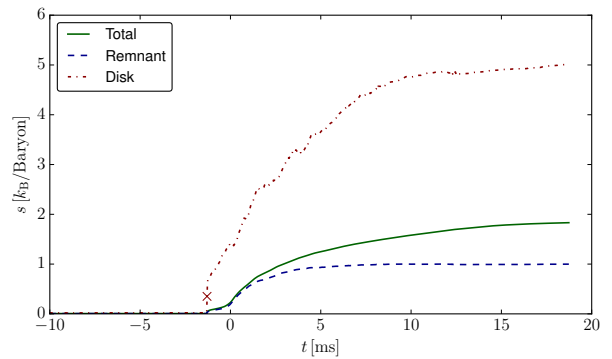

  \begin{center}
    \includegraphics[width=0.95\columnwidth]{{{entropy}}}  
    \caption{Evolution of average specific entropy for matter inside and outside
    the bulk mass, labeled tentatively as remnant and disk. The average specific 
    entropy is defined as the ratio of total entropy and total baryon number.
    The cross marks the activation of the thermal evolution.}
    \label{fig:entropy}
  \end{center}
\end{figure}

Next, we investigate the production of entropy during and after merger.
For this, we first divide the system into bulk and everything else. 
We then compute a specific bulk entropy from the total entropy in the bulk 
divided by the baryon mass of the bulk. In the same fashion, we compute
the specific entropy of the rest of the system, which we identify with
the disk (once the disk has formed). The result is shown in \Fref{fig:entropy}. The average specific 
entropy of the bulk reaches ${\approx} 1~k_B$ per baryon around $5\usk\milli\second$ 
after merger and then stays almost constant. This is consistent with the 
observation that no shock waves occur inside the bulk after this time.
The disk, on the other hand, is continuously heated by shocks waves originating
from the strongly deformed rotating outer envelope of the remnant.
Correspondingly, the specific entropy of the disk keeps increasing until the 
end of the simulation, albeit at a slowing rate. After $18\usk\milli\second$,
it reaches $5\usk k_B$ per baryon.

On a technical side note, we remark that the specific entropy of the bulk exterior jumps
abruptly to $0.5 \usk k_B$ shortly before merger when we switch from adiabatic 
evolution during inspiral to fully thermal evolution. At this point, there is 
no disk and the exterior of the bulk consists of the NS surface, which is 
notorious for numerical errors in hydrodynamic codes. The jump in the total 
entropy is insignificant compared to the final value. However, the heating at 
the NS surface due to numerical errors might have been more significant if it
had been allowed to accumulate during the whole inspiral. Since we observed no
adverse side effects from enforcing adiabatic evolution during inspiral, we 
recommend it for similar simulations.

\begin{figure}
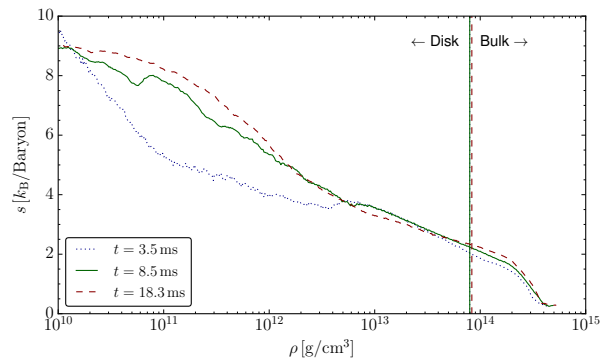

  \begin{center}
    \includegraphics[width=0.95\columnwidth]{{{sentr_vs_dens}}}  
    \caption{Average specific entropy of matter versus rest mass density.
    The time in the labels is the time after merger. The vertical lines
    mark the density defining the bulk.}
    \label{fig:sentr_vs_dens}
  \end{center}
\end{figure}

Using histograms summing the entropy and baryonic mass of numerical grid cells 
in (logarithmic) bins of rest mass density, we obtain the average specific entropy 
for matter at given density, defined as the entropy in the corresponding bin 
divided by the baryonic mass. The result is shown in \Fref{fig:sentr_vs_dens}.
We note that we do not consider neutrino radiation, which would likely reduce
the disk temperature. Also note that the temperature inside the remnant is not
just a function of density, as will be discussed in \Sref{sec:deformation}.

\subsection{Remnant rotation and mass profiles}  
\label{sec:remnant}

\begin{figure}
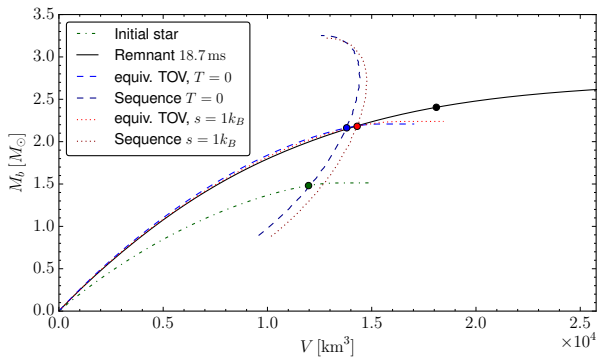

  \begin{center}
    \includegraphics[width=0.95\columnwidth]{{{mass_volume}}}  
    \caption{Mass distribution of the remnant $18.7\usk\milli\second$ after merger. 
    Plotted is the baryonic mass versus the proper volume contained in surfaces 
    of constant mass density. For comparison, we also show the same profile for 
    one of the initial stars in isolation, and for the TOV core equivalents (see 
    main text) with zero temperature as well as constant specific entropy of 
    $1\usk k_B$. The symbols mark, from left to right, the bulk (see main text) 
    of initial star, cold and hot core equivalent, and remnant. Further, we show 
    the bulk mass versus the bulk volume for the two sequences of TOV stars with 
    $T=0$ and $s=1\usk k_B$. }
    \label{fig:mass_volume}
  \end{center}
\end{figure}

We now discuss the density profile of the remnant near the end of the simulation, 
after it had time to settle down. As discussed in \Sref{sec:tools}, it is difficult 
to specify profiles in terms of radii unambiguously in GR in the absence of spherical
symmetry. Therefore, we compute the total baryonic mass and the proper volume inside
the isosurfaces of mass density, obtaining a mass-inside-volume relation that is
completely independent of the spatial coordinates.
\Fref{fig:mass_volume} shows the relation for the remnant. For comparison, we also
show this relation for one of the initial NSs at isolation.

Clearly, there is no well defined boundary between remnant and disk. To make quantitative 
statements about the remnant, we will use the definition of the bulk
as the isodensity surface with the maximum compactness (in the sense defined in
\Sref{sec:tools}). 
The position of the remnant bulk is marked in the figure as well,
and its main properties are listed in \Tref{tab:bulks}.
Figure~\ref{fig:mass_volume} also contains the relation of
bulk mass versus bulk volume for a sequence of TOV stars with the same EOS
as the initial data. Not surprising, the remnant bulk occupies more volume 
than a TOV star with the same bulk mass. 

Note that the curve corresponding to the TOV sequence intersects the 
mass-inside-volume relation of the remnant. The mass-inside-volume profile
for the TOV star at the intersection is also shown in \Fref{fig:mass_volume}. It is 
noteworthy that the profile of this TOV star and the remnant agree very well
inside the bulk of the TOV star. In the following we call the latter the 
``TOV core equivalent.'' Its properties are given in \Tref{tab:bulks}, which also
lists the central densities. Note the central density of the remnant and the TOV 
core equivalent agree within 3\%. We conclude that
the average density of the core is not affected significantly by rotation
and thermal effects. Note, however, that our measure is rather insensitive to 
nonradial deformations and cannot be used to draw conclusions on the oblateness
or non-radial oscillation amplitude of the core.

To estimate the influence of thermal effects on the core, we also computed the TOV sequence
and TOV core equivalent assuming beta equilibrium and a constant specific entropy
of $1\usk k_B$, which is the average value for the bulk of the remnant 
(cf.~\Fref{fig:entropy}). The bulk properties of the hot TOV core equivalents are 
listed in \Tref{tab:bulks}.
The hot TOV core equivalent indeed fits the remnant core better. However, as shown 
in \Fref{fig:mass_volume}, the differences to the cold star are rather small. 
The thermal pressure is therefore unimportant for the overall remnant core 
density in the case under consideration. As we will see in the next section, it is more 
relevant for nonaxisymmetric deformations of the core. Also the outer layers
are typically affected more strongly by thermal effects; e.g., the surface radius 
of the hot TOV core equivalent is considerably larger compared to the cold one.

\begin{table}
\caption{Bulk properties of remnant at $18.7\usk\milli\second$ after merger,
of one of the stars at the beginning of the simulation, and of the remnants 
TOV core equivalent. The latter is computed both for zero temperature (cold TOV) 
and for constant specific entropy of $1\usk k_B$ (hot TOV).
$M_\mathrm{blk}$ is the baryonic mass of the bulk, $V_\mathrm{blk}$ its proper 
volume, $\bar{s}_\mathrm{blk} = S_\mathrm{blk}/M_\mathrm{blk}$ its average 
specific entropy, and $S_\mathrm{blk}$ its total entropy. $\rho_c$ is the central
density.
}\label{tab:bulks}
\begin{tabular}{l|c|c|c|c}
         & $M_\mathrm{blk} \left[M_\odot\right]$ 
         & $V_\mathrm{blk} \left[10^{4} \usk\kilo\meter\cubed\right]$ 
         & $\bar{s}_\mathrm{blk} \left[k_B\right]$ 
         & $\rho_c \left[10^{14} \usk\gram\per\centi\meter\cubed\right]$\\ \hline
Remnant  & $2.405$  & $1.81$   &  $1.0$ & $5.30 $\\ 
Cold TOV & $2.163$  & $1.38$   &  $0$   & $5.47$ \\ 
Hot TOV  & $2.181$  & $1.43$   &  $1$   & $5.40$\\ 
Initial NS &$1.481$ & $1.20$   &  $0$   & $4.30$     
\end{tabular} 
\end{table}

\begin{figure}
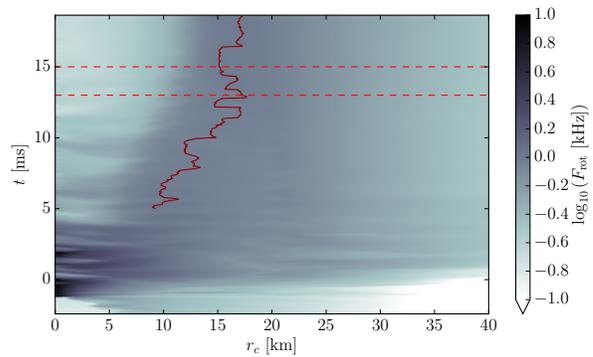

  \begin{center}
    \includegraphics[width=0.95\columnwidth]{{{omega_rt}}}  
    \caption{Evolution of $\phi$-averaged angular velocity in the equatorial plane.
    $r_c$ is the circumferential radius and $t$ the coordinate time relative to the
    merger time. The location of the maximum after the merger phase is marked by the 
    solid curve. The dashed horizontal lines mark the boundaries of the time-average 
    shown in \Fref{fig:rot_prof}.}
    \label{fig:omega_rt}
  \end{center}
\end{figure}

We now investigate the rotation profile of the remnant. To reduce gauge effects, 
we use the coordinate system in the orbital plane described in \Sref{sec:tools} 
to compute the average angular velocity (as seen from infinity) along coordinate circles.
The angular velocity as a function of time and circumferential radius is shown in 
\Fref{fig:omega_rt}. As one can see, directly after merger the average rotation rate
in the core is highest. Note, however, that during this phase the system is not even 
approximately axisymmetric or stationary. In any case, the fluid flow quickly 
rearranges itself, and $4\usk\milli\second$ after merger the core rotates slower
than the parts at larger radii. 

During the rearrangement phase, the development of a Kelvin-Helmholtz instability 
is expected. We note that the 
numerical resolution is too low to resolve it in full. This might have an effect 
on the final rotation profile, which is, however, difficult to quantify without
very high resolution simulations.

\begin{figure}
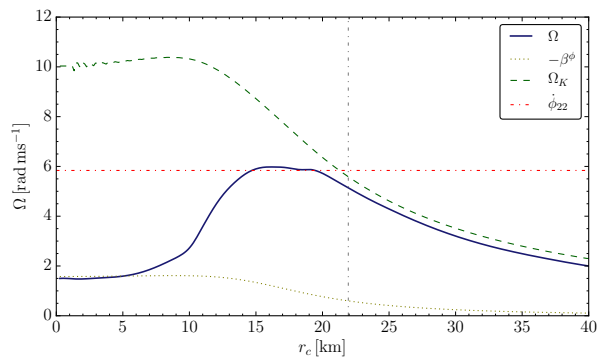

  \begin{center}
    \includegraphics[width=0.95\columnwidth]{{{rot_prof}}}  
    \caption{Rotation profile of the remnant $14\usk\milli\second$ after merger.
    Shown is the angular velocity $\Omega$ in the equatorial plane averaged in $\phi$ direction as
    well as in time over a $2\usk\milli\second$ window ($13-15\usk\milli\second$). For comparison, the green dashed line
    is the orbital angular velocity $\Omega_K$ of a test particle in corotating circular orbit. 
    The dotted grey line is the frame-dragging contribution.
    The horizontal line is the pattern angular velocity $\dot{\Phi}_{22} = \dot{\phi}/2$ of 
    the $l=m=2$ component of the gravitational signal at the same (retarded) time.
    The vertical line marks the radius where the density falls below 5\% of the maximum one.}
    \label{fig:rot_prof}
  \end{center}
\end{figure}

The rotation profile after the remnant has settled down is shown in \Fref{fig:rot_prof}.
To minimize the influence of remaining oscillations, we averaged in time between
$13$--$15 \usk \milli\second$. We find that the maximum angular velocity is around 
$6\usk\radian\per\milli\second$
and occurs at radii in the range $15$--$20\usk\kilo\meter$.
In order to compare the rotation rate to the Kepler velocity, we also averaged the spacetime
in time and $\phi$-direction, and we computed the angular velocities of test masses
in corotating circular orbits, as described in \cite{Kastaun:2015:064027}.
The result is shown in \Fref{fig:rot_prof}, as well as the contribution of frame dragging
to the angular velocity. We find that the core rotates indeed much slower
than the local Kepler velocity, and is basically nonrotating in the local inertial frame.
Further, we find that the outer layers approach the Kepler velocity.

Note that similar rotation profiles have already been observed in \cite{Kastaun:2015:064027,Endrizzi:2016:164001} 
for other binary models with different EOS and masses resulting in hyper- and 
supramassive remnants.
Figure~\ref{fig:rot_prof} also exhibits a feature not yet explained: the maximum rotation rate 
is very close to the angular velocity of the $l=m=2$ component of the GW pattern.
This too has been observed in \cite{Kastaun:2015:064027,Endrizzi:2016:164001} and thus
seems more than a coincidence.

\subsection{Remnant deformation and thermal structure}
\label{sec:deformation}

In this section, we will describe the evolution of the non-axisymmetric density perturbation 
of the remnant, the hot spots, the fluid flow, and their relation to each other.

\begin{figure}
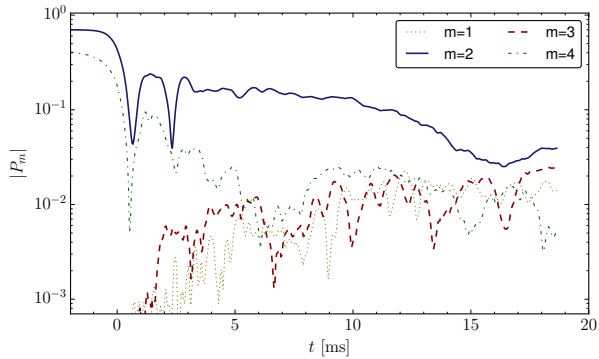

  \begin{center}
    \includegraphics[width=0.95\columnwidth]{{{moments_amp}}}  
    \caption{Evolution of mass density moments in the equatorial plane.}
    \label{fig:moments_amp}
  \end{center}
\end{figure}

The simplest quantitative measure for the deformation of the remnant is the decomposition 
of the density in the equatorial plane into $\phi$-harmonic components, as described in \Sref{sec:tools}. 
The moments $P_m^\rho$ up to $m=4$ are shown in \Fref{fig:moments_amp} as a function
of time. Clearly, the merger excites $m=2$ and $m=4$ deformations (the latter probably being a 
nonlinear overtone of the former). Since we study an equal mass system, no odd components
are excited at merger. During the evolution, however, we observe a growing $m=3$ and $m=1$
contribution which becomes comparable to the $m=2$ component at the end of the simulation.
The mechanism of this instability is not explained yet. The occurrence of $m=3$ modes 
for other equal mass systems has already been reported in \cite{Bernuzzi:2014:104021}.

\begin{figure*}
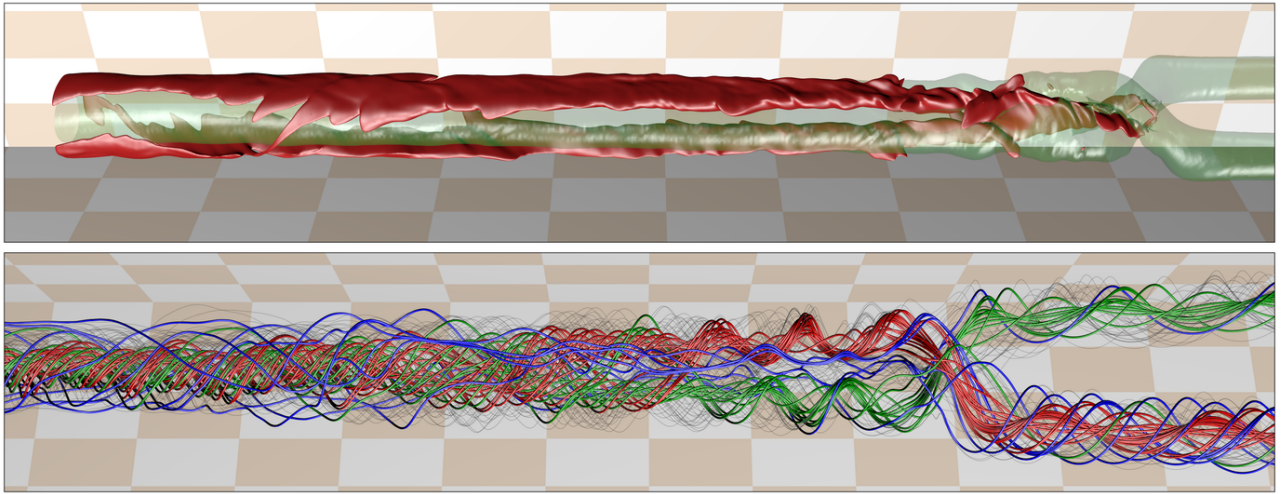

  \begin{center}
    \includegraphics[width=0.95\textwidth]{{{sht_s1_unscrewed_small}}}  
    \caption{Spacetime diagram showing the fluid evolution in the orbital plane.
    The time coordinate runs from right to left. Each time slice has been
    rotated to the phase of the $m=2$ component of the density perturbation
    in the orbital plane. The size of the tiles in the background is 
    $2\usk\milli\second \times 20 \usk\kilo\meter$.
    Top panel: the red solid surface is the contour of constant entropy density 
    chosen to highlight the evolution of the hot spots, and the green transparent
    surface is the world-tube of the isodensity contour containing (at each time) 
    25\% of the total baryon mass.
    Bottom panel: traced trajectories of fluid elements. Only elements not ejected from 
    the remnant are shown. The colors highlight groups of trajectories discussed in 
    the main text.
    }
    \label{fig:sht_s1_unscrewed}
  \end{center}
\end{figure*}

We now turn to investigate the nature of the hot spots. As can be seen in the lower panel of
\Fref{fig:sht_s1_screwed}, the hot spots seem to be phase locked with the density perturbation.
To make this even more visible, we produced a similar spacetime diagram in which we rotated
each time slice by half of the complex phase of the $m=2$ component of the density perturbation 
in the equatorial plane. In the following, we simply call those coordinates
corotating. The result is shown in \Fref{fig:sht_s1_unscrewed}.
The upper panel clearly demonstrates that the hot spots and the density perturbation are in 
phase with each other for most of the evolution.
Only near the end, the picture is complicated by the decreasing amplitude of the $m=2$
mode and the growing contribution of the $m=3$ mode.
Note that the sudden changes of the entropy density isocontour in the late phase of the run
are not caused by strong fluid movements. On the contrary, the entropy becomes more axisymmetric
and the location of the isocontour more sensitive to small fluctuations.   

The stability of the hot spots in the post merger phase raises the question why those 
are not dissipated by the differential rotation (cf.~\Fref{fig:rot_prof}).
To address this question, we investigate the fluid flow in relation to the hot 
spots. To get an overview, we traced fluid elements, as described in \Sref{sec:tools}, 
and transformed their trajectories in the corotating coordinate system.  The result is shown 
in the lower panel of \Fref{fig:sht_s1_unscrewed}. We find that after the merger,
there are initially two large vortices with the same orientation, and a shear layer in between.
For better visibility, we colored some trajectories belonging to one of the two vortices red and green, 
respectively. Over time, the two vortices merge into a larger one which can be interpreted as simple
differential rotation on top of a large $m=2$ deformation. Note that each of the two vortices 
consists mainly of matter originating from one particular star, leading to the interpretation 
that the cores remain more or less independent for a few milliseconds.

We note that the shear layer is subject to the Kelvin-Helmholtz (KH) instability, which is 
severely under-resolved at the resolutions used in our simulation. For a discussion of the 
KH instability in BNS mergers we refer to \cite{Kiuchi:2015}, presenting results employing the 
finest grid spacing to date.
Here we just caution that a correct evolution of the KH mechanism will most likely change
the timescale on which the two vortices merge, although we believe that the qualitative picture
will remain unchanged.

\begin{figure}
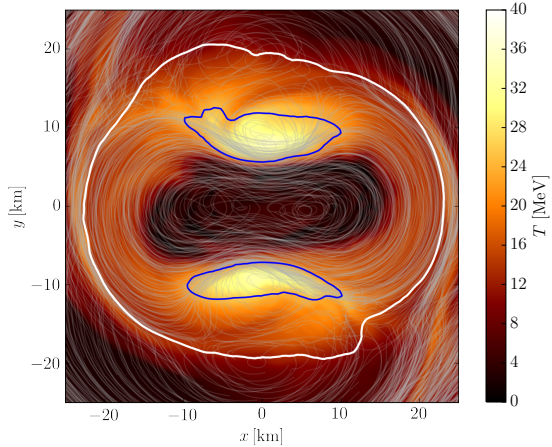

  \begin{center}
    \includegraphics[width=0.95\columnwidth]{{{temp_traj_xy}}}  
    \caption{Fluid motion in the frame corotating with the $m=2$ component of the density
    perturbation. The color plot shows the temperature in the orbital plane 
    $8.5\usk\milli\second$ after merger. The curves show fluid trajectories during the 
    time interval $\pm 2 \usk \milli\second$ around the snapshot.
    The solid  blue line marks the entropy density isocontour whose worldtube is shown 
    also in \Fref{fig:sht_s1_unscrewed}, the thick white line marks the bulk isodensity 
    surface.}
    \label{fig:temp_traj_xy}
  \end{center}
\end{figure}

\begin{figure}
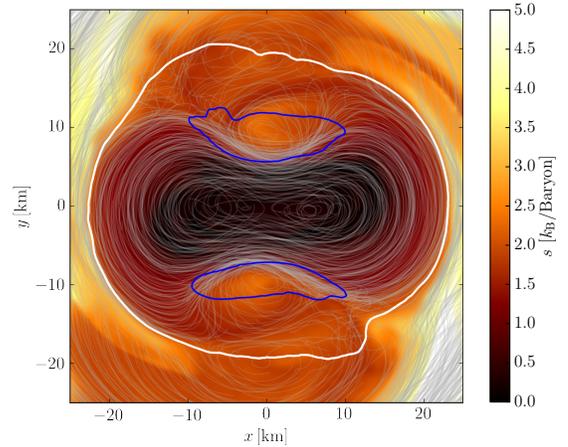

  \begin{center}
    \includegraphics[width=0.95\columnwidth]{{{sentr_traj_xy}}}  
    \caption{Like \Fref{fig:temp_traj_xy}, but showing the specific entropy.}
    \label{fig:sentr_traj_xy}
  \end{center}
\end{figure}

We also found a group of fluid elements (colored blue in the figure) which move only slowly 
relative to the $m=2$ density perturbation. Based on \Fref{fig:sht_s1_unscrewed} we find
that those trajectories are roughly in the vicinity of the hot spots. It is important
to know if the quasi-stationary trajectories agree exactly with the location of the hot 
spots, which would explain their stationarity with respect to the $m=2$ density perturbation.
To answer this, we created another plot showing the temperature and specific entropy distribution
in the equatorial plane in \Fref{fig:temp_traj_xy}. On top of this, we plot the trajectories in 
the coordinates corotating with the $m=2$ perturbation, over a brief period ($4\usk\milli\second$) around the time 
of the snapshot. We find that part of the high-temperature regions inside the remnant are indeed
advected along the quasi-stationary fluid trajectories which form a small vortex.
However, many fluid trajectories just pass through the hot spots, with the corresponding 
fluid elements periodically heating up and cooling down. Further, the trajectories also converge 
where they approach the hot spots (at least in the orbital plane). This means that
part of the hot spots are dynamic in nature, generated by adiabatic temperature changes along
a locally compressive flow. As further evidence, we plot the specific entropy for the same time,
as shown in \Fref{fig:sentr_traj_xy}. The specific entropy in the plot varies only slightly 
along the trajectories crossing the hot spots. Of course, plotting trajectories on top 
of a snapshot showing one instant of time is only meaningful to the degree to which the 
thermal pattern remains stationary. By comparing to animations of the same cuts, we find
that, although smaller thermal features are advected along the flow, the main hot spots 
are indeed largely due to the local adiabatic compression. The hot matter trapped initially 
in the small vortices is redistributed toward the end of the 
simulation and becomes part of the ring-shaped structure shown in the lower right panel
of \Fref{fig:entropy_3d_snaps}.

\subsection{Gravitational waves}
\label{sec:gw}
In this section we present the GW signal, with focus on the postmerger signal. 
We are mainly interested in relating the qualitative features to the remnant 
dynamics. 
In \cite{Endrizzi:2016:164001}, a resolution study for a different model but 
similar number of grid points per initial stellar radius was carried out, but 
could not demonstrate convergence for the complete postmerger phase. This makes
it difficult to provide error estimates.

To compute the GW signal, we use an extraction radius of $916 \usk\kilo\meter$ 
and do not extrapolate to infinity.
The time integration required to obtain the strain has been carried out using the 
fixed-frequency integration described in \cite{Reisswig:2011:195015}, with a 
cutoff frequency of $500\usk\hertz$.
At the resolution of the coarsest grid, a wave with $3\usk\kilo\hertz$ is resolved 
by 10 grid points. Higher frequencies would be increasingly suppressed. However, we 
compared the signals extracted inside finer grids to verify that no significant 
high frequency signals are generated.

\begin{figure}
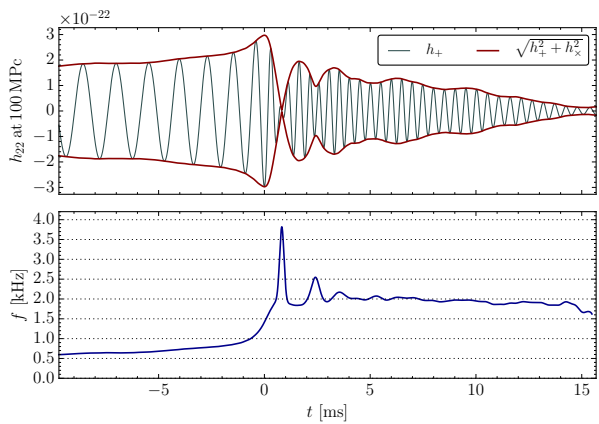

  \begin{center}
    \includegraphics[width=0.95\columnwidth]{{{gw_strain}}}  
    \caption{Gravitational wave strain at $100\usk\mega\parsec$ (top panel) 
    and instantaneous frequency (bottom panel).
    The latter is computed from the phase velocity of the complex strain, smoothed over 
    $0.1\usk\milli\second$ to avoid amplification of high-frequency noise.}
    \label{fig:gw_strain}
  \end{center}
\end{figure}

We find a post merger signal dominated by the $m=2$ perturbation of the remnant.
The evolution of the strain amplitude is shown in \Fref{fig:gw_strain}, in terms
of the coefficient $h_{22}$ in the expansion in spin-weighted harmonics (see
\cite{Thorne:1980:299}). The strain observed at a given viewing angle $\theta$
can be obtained by multiplying with the norm of the spin weighted harmonic, 
$|{}_{-2}Y_{22}(\theta, \phi)|$. We also plot the instantaneous frequency
extracted from the phase of the complex strain. 
As one can see, the strain amplitude decays significantly 
within $15\usk\milli\second$, as can be expected from the decrease of the $m=2$
fluid perturbation shown in \Fref{fig:moments_amp}.
The frequency shows some mild oscillations after merger and then remains stable
in the range $1.8$-$2\usk\kilo\hertz$, apart from a small drift. We note that the 
first peak in the instantaneous frequency is not very meaningful since the strain 
amplitude at this point has a rapid zero crossing. 

\begin{figure}
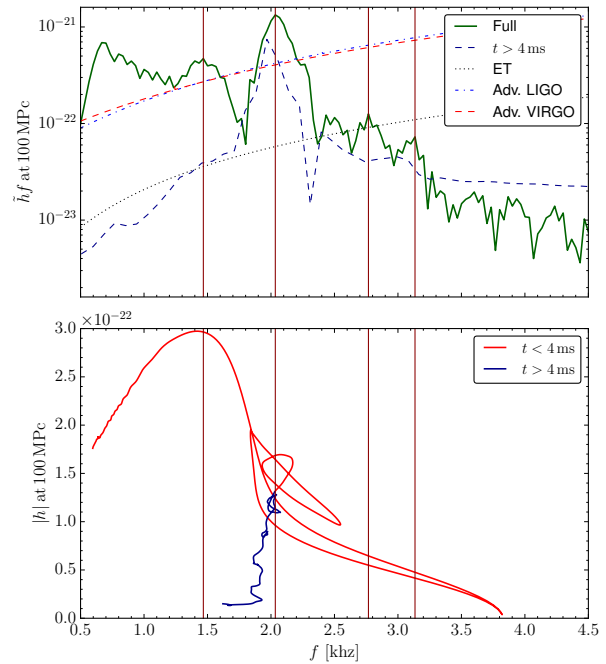

  \begin{center}
    \includegraphics[width=0.95\columnwidth]{{{gw_amp_freq}}}  
    \caption{Top panel: gravitational wave spectrum at $100\usk\mega\parsec$, 
    both for the full simulation (solid green line) and for the strain starting 
    $4\usk\milli\second$ after the merger (dotted blue line). For comparison, we
    show the sensitivity curves of GW detectors.
    Bottom panel: strain amplitude versus instantaneous frequency.
    The part starting $4\usk\milli\second$ after the merger is plotted in blue,
    the earlier evolution in red.}
    \label{fig:gw_amp_freq}
  \end{center}
\end{figure}

The lower panel of \Fref{fig:gw_amp_freq} shows the evolution of the GW signal 
in a different way, plotting the strain amplitude versus instantaneous frequency.
Both frequency and amplitude are growing during inspiral, reaching a peak amplitude
at a frequency $1.4\usk\kilo\hertz$. During the merger, the frequency keeps
increasing due to the increased rotation rate. At the same time, the quadrupole
moment decreases as the system becomes more compact, resulting in a net decrease
of strain amplitude. Next, the forming remnant reaches maximum compression,
reflected in a high rotation rate.  Apparently the quadrupole moment undergoes a zero 
crossing during this period, resulting in the brief quenching of the GW strain.
Next, the remnant expands again, entering the double core phase discussed earlier.
The rotation rate and thus the frequency of the GW signal decrease, while the amplitude
increases because of the increasing quadrupole moment.
The system then experiences a milder bounce, with an increase of frequency up to 
$2.5\usk\kilo\hertz$. This time the amplitude is only reduced, but stays clearly
above the late-time amplitude, making this phase a significant part of the postmerger 
signal. Finally, the star settles down and the amplitude decreases continuously,
while the frequency only shows a small drift.  This late part is colored blue in the 
plot.

The corresponding GW power spectrum is shown in the top panel of \Fref{fig:gw_amp_freq}
in comparison to the design sensitivities of current and future GW detectors.
Besides the inspiral and plunge signatures, the post-merger evolution causes a
prominent peak around $2\usk\kilo\hertz$. Comparing to the sensitivity curves, 
we find that the merger signal should be detectable by advanced LIGO/Virgo
up to a distance $10-100\usk\mega\parsec$. We also find that if the merger itself
can be detected, the post-merger signal should be visible as well. 
To distinguish contributions from early post-merger phase and later evolution,
we computed the spectrum of the signal starting $4\usk\milli\second$ after
merger. We find, as expected from the instantaneous frequency, a slightly
smaller peak frequency, although the difference is less than the width of the peak.  
More importantly, the amplitude of the peak is reduced
by a factor of 2, meaning that the evolution directly after merger is an important
part of the post-merger signal. 
We note, however, that our remnant is stable and might potentially radiate for a long time,
which we did not consider in our spectra.

\subsection{Disk and matter ejection}
\label{sec:disk}

\begin{figure}
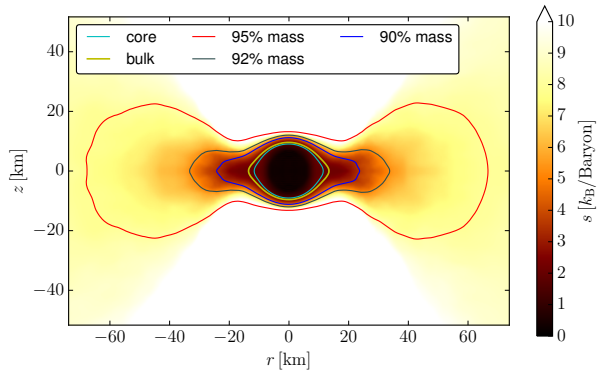

  \begin{center}
    \includegraphics[width=0.99\columnwidth]{{{disk_sentr_xz}}}  
    \caption{Disk structure at the end of the simulation, averaged in time
    over $4\usk\milli\second$. The colors correspond
    to the specific entropy, while the lines are the isocontours of mass density 
    $1.43\times 10^{14}$, $8.43\times 10^{13}$, 
    $5.43\times 10^{12}$, $1.20\times 10^{12}$, and
    $1.25\times 10^{11}\usk \gram\per\centi\meter\cubed$, from innermost to outermost, 
    respectively. The total mass of matter with higher density than the plotted density 
    levels is equal to the bulk mass of the TOV core equivalent,
    the bulk mass of the remnant itself, 90\%, 92\%, and 95\% of the total baryon mass, 
    respectively.
     }
    \label{fig:disk_sentr_xz}
  \end{center}
\end{figure}

In the following, we discuss the properties of the disk surrounding the remnant.
The structure in the meridional plane is shown in \Fref{fig:disk_sentr_xz}.
Since the rotation rate of the remnant (compare \Fref{fig:rot_prof}) approaches 
the Kepler rate in the outer layers, the transition between the disk and the remnant 
is smooth. The plot also shows the specific entropy, which is increasing
with radius, and also in the vertical direction. This agrees with 
\Fref{fig:sentr_vs_dens}, where we saw an increase of specific entropy with decreasing 
mass density. 
Figure~\ref{fig:disk_sentr_xz} also shows the shape of the bulk of the remnant
defined in \Sref{sec:tools}, demonstrating again that using 
the bulk properties to describe the remnant is meaningful.

\begin{figure}
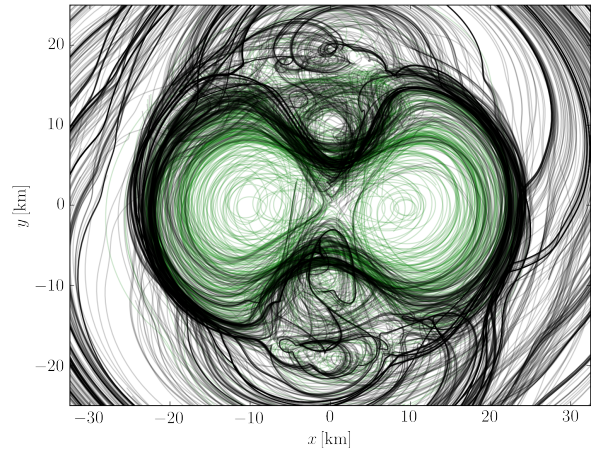

  \begin{center}
    \includegraphics[width=0.95\columnwidth]{{{traj_xy}}}  
    \caption{Fluid trajectories in the orbital plane with respect to the 
    coordinate system corotating with the $m=2$ density perturbation. The
    trajectories are shown for the time interval $2$ to $12\usk\milli\second$
    after merger. The trajectories have been tracked backwards in time starting 
    $14 \usk\milli\second$ after merger from seeds arranged in a regular grid with
    extend $70 \usk\kilo\meter$. The trajectories are colored green if they stay
    in the remnant until the end of the simulation, else black. Note the 
    trajectories move predominantly clockwise.
     }
    \label{fig:traj_xy}
  \end{center}
\end{figure}

Besides the structure of the disk, we also investigated its formation.
To this end, we traced fluid elements backwards in time, starting from a regular grid
in the orbital plane covering the disk at the end of the simulation. We then plotted the 
trajectories in the coordinate system corotating with the phase of the $m=2$ component 
of the density perturbation of the remnant. The resulting trajectories, starting 
$2\usk\milli\second$ after merger, are shown in \Fref{fig:traj_xy}. 
Fluid elements that end up in the disk at the end of the simulation are colored
black, while those remaining in the remnant are colored green. Since
the trajectories are traced backwards in time starting with a homogeneous tracer
density, the dense bundles of trajectories visible in the plot correspond to a diverging
fluid flow, which is to be expected assuming that the density of a fluid element 
in the disk is lower compared to its point of origin inside the merging stars.

From \Fref{fig:traj_xy}, we can see that most of the matter in the disk at some point came
close to the outermost parts of the strongly deformed remnant, which is not surprising.
More interesting is the fact that many trajectories follow a bean-shaped trajectory along the 
outer layers of the remnant before migrating to the disk. Also, the little vortices
tailing behind the elevated parts of the deformed remnant 
seem to be involved in the migration of matter into the disk. It is also worth noting that 
apparently some of the trajectories undergo radial bounces before escaping into the disk, 
probably related to the radial oscillations of the remnant.

Finally, we studied the matter ejection using the methods described in \Sref{sec:tools}.
The amount of ejecta is relatively small, ${\approx} 3\times 10^{-4}\usk M_\odot$, and is 
emitted mainly in a single wave a few milliseconds after merger.
The average electron fraction is around $0.05$. We note that this reflects the electron 
fraction from where the ejected matter originates during the merger. Considering the 
neutrino radiation would most likely change the final electron fraction of the ejecta 
considerably (compare \cite{Radice:2016:1108}).
The average specific entropy of the ejected matter is ${\approx}15\usk k_\mathrm{B}$ per baryon,
and the escape velocity ${\approx}0.12\usk c$.
We note that a binary with the same EOS but a higher total baryonic mass of $4 \usk M_\odot$ 
was evolved in \cite{Kastaun:2015:064027}, and the resulting ejecta mass was around 20 times 
larger. This might be related to the fact that the heavier model showed stronger radial 
oscillations after merger.

\section{Summary and Conclusions}
\label{sec:summary}

In this work, we investigated the merger of two NSs, each with a
mass of $1.4 \usk M_\odot$, employing the Shen-Horowitz-Teige EOS.
Because of the atypically large maximum mass allowed by this EOS, the 
result is a stable NS. 
We followed in detail the fluid flow during and after merger and
found that for around $8$-$10\usk\milli\second$, the stellar cores
remain roughly independent, rotating against each other while orbiting
around each other. For the first ${\approx}3\usk\milli\second$, they are 
separated by a shock-heated layer with lower density. This layer is quickly 
ripped apart by the fluid flow, but the cores start merging only when the 
shear layer in between is dissipated into smaller vortices, with an 
increasing number of fluid trajectories enclosing both cores.
Although our resolution is not sufficient to fully resolve the 
Kelvin-Helmholtz instability, we believe the qualitative picture is 
correct (even if the timescales might change).
After the cores have merged, the fluid flow in the equatorial plane 
is peanut-shaped, following a strong $m=2$ perturbation.

We studied the structure of the remnant near the end of the simulation, after 
it has settled down. The rotation profile has a maximum at a radius 
$15$--$20\usk\kilo\meter$, reaching a rotation rate around $960\usk\hertz$, 
while the core is slowly rotating with a rate of only $240\usk\hertz$ measured 
from infinity.
Moreover, this rate is mostly due to frame dragging, and the core is almost
nonrotating in the local inertial frame. The remnant has extended outer layers 
which approach Kepler velocity and smoothly join the disk orbiting the remnant.

An important open question is about the prospect for the formation of a disk surrounding 
the BH even if the collapse happens much later than the merger, so that the original
disk is already accreted onto the SMNS. This has only been investigated
for simplified models with $j$-const rotation law or uniform rotation
\cite{Baiotti:2007:187, Giacomazzo2011PhRvD..84b4022G, Margalit2015}.
In this respect, a rotation profile like the one described above might or might 
not turn out to provide the necessary conditions. Of course, this would also require
that the rapid rotation of the outer bulge of the remnant persists longer than the disk.
This aspect will be further investigated in future work, and in particular in relation 
to the ``time-reversal" scenario for SGRBs \cite{Ciolfi:2015:36}, which envisages 
a late time collapse of the remnant into a BH plus accretion disk. 

To study the mass distribution of the remnant, we introduced a new measure
that replaces density profiles, mass, and compactness in a way that can be used
unambiguously for rapidly and differentially rotating merger remnants without a
clearly defined surface. We found that the core has a structure very similar 
to the core of a specific TOV star. We set up both a cold TOV solution as well 
as one with the same average specific entropy as the remnant, and found that,
although the hot star fitted the remnant better, thermal pressure has only a
little effect on the average density profile of the core.

Investigating the thermal evolution in more detail, we found that the average specific 
entropy of the remnant reaches $1 \usk k_B$ at $5\usk \milli\second$ after merger and 
then stays constant, while the disk is continuously heated up, reaching an average
specific entropy $5\usk k_B$ at the end of the simulation.
At this point, the average specific entropy at a given density raises from 
$0.3 \usk k_B$ at the center of the remnant to around $3 \usk k_B$ at the 
transition to the disk, and continues increasing toward lower densities. 
We note, however, that our simulation does not include neutrino radiation, which 
would likely change the temperature of the disk on the timescale of our evolution.

One of our main interests was on the inhomogeneity of the remnant temperature.
We found that quickly after merger the entropy of the remnant is concentrated
in two hot spots which remain stable for around $15\usk\milli\second$.
Those hot spots are locked in phase with the dominant $m=2$ perturbation, and
consist partially of hot matter trapped in a small vortex in the ``waist'' of the
peanut shaped remnant deformation. A significant part is however of dynamic nature,
caused by adiabatic heating along a local compression. Because of the additional
thermal pressure, it seems likely that the dynamic hot spot in turn has a 
backreaction on the remnant deformation and hence the fluid flow. 
The hot spots can thus alter the amplitude of the GW signal caused by the 
main deformation, but, being phase locked, do not contribute additional peaks to
the power spectrum for the case at hand. We note, however, that in general secondary
vortices such as the ones in our model might also undergo instabilities. Our study
shows that the remnant structure is clearly affected by the vortices/hot spots; hence 
a rearrangement might slightly alter the remnant structure and hence lead to a sudden 
change of the GW frequency. 

The quantitative results we find are not general. Almost certainly results will change 
when considering unequal mass systems. Also the initial spin of the NSs has a strong
influence, as we will show in a forthcoming publication.
Nevertheless, we can conclude that the deformation of the remnant up to at 
least $20\usk\milli\second$ after merger should be treated as a complex nonlinear 
quasi-stationary state with relevant dynamic thermal effects. 
A state such as the one found in our work cannot be described well in 
terms of linear perturbations of axisymmetric rotating stars.
Since the amplitude of the postmerger GW signal is decaying, this complicated 
period is highly relevant for the interpretation of future detections.  
Interpreting minor features of the GW spectrum such as side peaks in terms of 
oscillation mode analysis developed for isolated stars might be misleading.
If a post-merger GW signal is detected, its duration can be taken as a lower limit 
for the lifetime of the remnant. In this 
regard it is relevant that a mounting number of results points 
to slowly rotating cores, even for HMNSs and SMNSs. The lifetime is then determined 
by the angular momentum balance of the outer layers and disk, not the core.

\acknowledgments
We acknowledge support from MIUR FIR grant No.~RBFR13QJYF.
The computations in this work have been carried out on the \texttt{Fermi}
cluster (CINECA) using a ISCRA class B allocation (IsB11).

\bibliographystyle{apsrev4-1-noeprint}
\bibliography{trento}

\end{document}